\documentclass[a4paper,11pt]{article}
\usepackage{pos}
\usepackage{graphicx}

\title{Modeling the Evolution from Massive Stars to Supernovae and Supernova Remnants}
\ShortTitle{Modeling the Evolution from Massive Stars to SNe and SNRs}

\author*[a]{Salvatore Orlando}

\affiliation[a]{INAF - Osservatorio Astronomico di Palermo, Piazza
del Parlamento 1, 90134 Palermo, Italy}

\emailAdd{salvatore.orlando@inaf.it}

\abstract{
The study of core-collapse supernova remnants (SNRs) presents a
fascinating puzzle, with intricate morphologies and a non-uniform
distribution of stellar debris. Particularly, young remnants (aged
less than 5000 years) hold immense value as they can offer crucial
insights into the inner processes of the supernova (SN) engine,
revealing details about nucleosynthetic yields and large-scale
asymmetries arising from the early stages of the explosion.
Furthermore, these remnants also bear characteristics that may
reflect the nature of their progenitor stars and the interactions
between the remnants and the surrounding circumstellar medium (CSM),
shaped by the progenitor's mass-loss history. Hence, investigating
the connection between young SNRs, parent SNe, and progenitor massive
stars can be of paramount importance to delve into the physics of
SN engines, and to investigate the final stages of massive star
evolution and the elusive mechanisms governing their mass loss. In
this contribution, I review recent advances in modeling the path
from massive stars to SNe and SNRs achieved by our team. The focus
is on investigating the links between the observed physical and
chemical properties of SNRs and their progenitor stars and SN
explosions. The unraveling of this connection offers us the opportunity
to probe the physics of core-collapse SN explosions and the final
stages of evolution of massive stars.}

\FullConference{
Multifrequency Behaviour of High Energy Cosmic Sources XIV (MULTIF2023)\\
12-17 June 2023\\
Palermo, Italy\\}

\begin{document}
\maketitle

\section{Introduction}

Core-collapse (CC) supernovae (SNe), the final fate of massive stars
(with masses larger than 8 M$_{\odot}$), play a crucial role in the
dynamic and chemical evolution of galaxies. They drive the chemical
enrichment of the diffuse interstellar gas and inject mass and
energy into the galaxies, including kinetic/thermal energy, cosmic
rays, and neutrinos. However, understanding the physical processes
governing SNe is a challenging task, requiring demanding three-dimensional
(3D) models to accurately follow the blast wave evolution. Moreover,
SNe are rare in our galaxy, and there have been no observed events
in the past four centuries. Consequently, our ability to study them
is confined to extragalactic SNe, which remain spatially unresolved
due to their substantial distance from us. Regrettably, obtaining
crucial insights into the explosion processes and the nature of the
progenitor systems from these observations presents significant
challenges. This is unfortunate because the period immediately
following a SN event, spanning tens to hundreds days, is when
the rapidly expanding debris from the stellar explosion retains a
full memory of the parent SN and progenitor star.

Nonetheless, the fingerprints of the explosion mechanisms and
progenitor systems may still be found in the outcome of SN explosions,
the supernova remnants (SNRs), hundreds to thousands of years after
the explosion (e.g., \cite{2021A&A...645A..66O}). These remnants appear
as extended sources emitting both thermal and nonthermal radiation
across different spectral bands, and they are also sources of cosmic
rays. By employing spatially resolved spectroscopy, astronomers
have been able to investigate the structure of nearby SNRs and the
distribution of chemical elements within them. This reveals a
complexity in their morphology that cannot be observed in unresolved
extragalactic sources \cite{2017hsn..book.2211M}.

The vast and high complexity observed in the remnants of CC-SNe is
believed to originate in part from pristine structures of the parent
SN explosion. These structures likely result from the
development of hydrodynamic/magnetohydrodynamic (HD/MHD) instabilities during
the launch of the anisotropic blast wave \cite{2013A&A...552A.126W,
2015A&A...577A..48W, 2017ApJ...842...13W, 2017hsn..book.1095J}. A
compelling example illustrating this phenomenon is the SNR Cassiopeia
A (Cas~A). The observed morphology and expansion rate of Cas~A
suggest that the remnant is still expanding through the nearly
spherically symmetric wind from its progenitor star. Consequently,
the prominent large-scale anisotropies observed in its morphology
are likely a result of processes associated with the SN explosion
\cite{2016ApJ...822...22O, 2021A&A...645A..66O, 2017ApJ...842...13W}.
In essence, these asymmetries in SNRs provide a unique opportunity
to delve into the physics of SN engines by gaining insight into the
processes that occur during the explosive event.

Moreover, the morphology of the remnants can yield crucial insights
into the structure of the stellar progenitor at the moment of
collapse, a stage that might have undergone significant alterations
due to complex and poorly-understood phenomena occurring during the
terminal phases of stellar evolution \cite{2014ApJ...785...82S,
2021ApJ...921...28F, 2021ApJ...908...44Y}. Recent research has
revealed that the geometric and physical properties of post-explosion
anisotropies are intricately linked to the density structure of the
progenitor star at collapse (e.g., \cite{2015A&A...577A..48W}). Since these
post-explosion anisotropies are expected to shape the structure of
the ejected material and ultimately influence the remnant's morphology,
the latter may harbor valuable information to unlock the nature of
the progenitor stars.

Lastly, certain SNR asymmetries arise from the dynamic interaction
between the remnant and an inhomogeneous circumstellar medium (CSM)
or interstellar medium (ISM). Notably, very young SNRs like the
remnant of SN 1987A showcase this phenomenon, where the highly
inhomogeneous CSM gives rise to the striking ring-like structure
observed across various wavelengths (e.g., \cite{2007Sci...315.1103M}).
Similarly, in more evolved SNRs like the Cygnus Loop, clear
indentations in the remnant outline attest to the effect of the
SNR's interaction with isolated dense interstellar clouds (e.g.,
\cite{2015SSRv..188..187S}). In young SNRs, the surrounding environment
is the CSM sculpted by the winds of the progenitor stars. Thus,
analyzing these asymmetries offers valuable insights into the mass
loss history of the progenitor stars and sheds light on the late
phases of their evolution.

In light of the above considerations, it becomes evident that
observations of SNRs can be very rich sources of information about
the parent SNe and the progenitor massive stars. By deciphering the
intricate features of these remnants, we can significantly enhance
our understanding of the complex evolutionary processes that
culminate in CC-SNe, thus unraveling the journey of massive stars
toward their ultimate fate. This knowledge is particularly importance
as it unravels the complexities inherent in stellar life cycles and
sheds light on the profound impact SNe can have on the ISM
and the dynamics and energetics of the Galaxy.

In the subsequent sections, I will review recent advances achieved
by our team in studying the ``progenitor-supernova-remnant connection''.
This consists in exploring how the physical, chemical, and morphological
properties observed in SNRs reflect the physics of SN engines and
the nature of the progenitor stellar systems.

\section{Modeling the path from massive stars to supernovae and
supernova remnants}
\label{method}

Combining the complementary information extracted from
multi-band/multi-messenger observations of candidate progenitor
stars, extragalactic SNe, and nearby SNRs can be essential to study
the processes associated to SNe and the latest stages of stellar
evolution. However, linking the asymmetries of the SNR bulk ejecta
with the information extracted from the analysis of SNe and with
the inferred information of the progenitor stars is a daunting task.
This requires a multi-scale, multi-physics, and multi-dimensional
(multi-D) approach that encompasses various essential aspects:
accounting for vastly different temporal and spatial scales involved
throughout the diverse phases of evolution; describing the structure
and chemical stratification of the progenitor star at the time of
collapse; understanding the complexities of explosive nucleosynthetic
processes; exploring the effects of inherently 3D post-explosion
anisotropies; investigating the interaction of SNe and SNRs with
the (magnetized) inhomogeneous CSM or ISM; incorporating the
back-reaction of accelerated cosmic rays on the shock dynamics;
synthesizing emission across different spectral bands to effectively
compare model results with observations.

\begin{figure}
  \begin{center}
    \includegraphics[width=15cm]{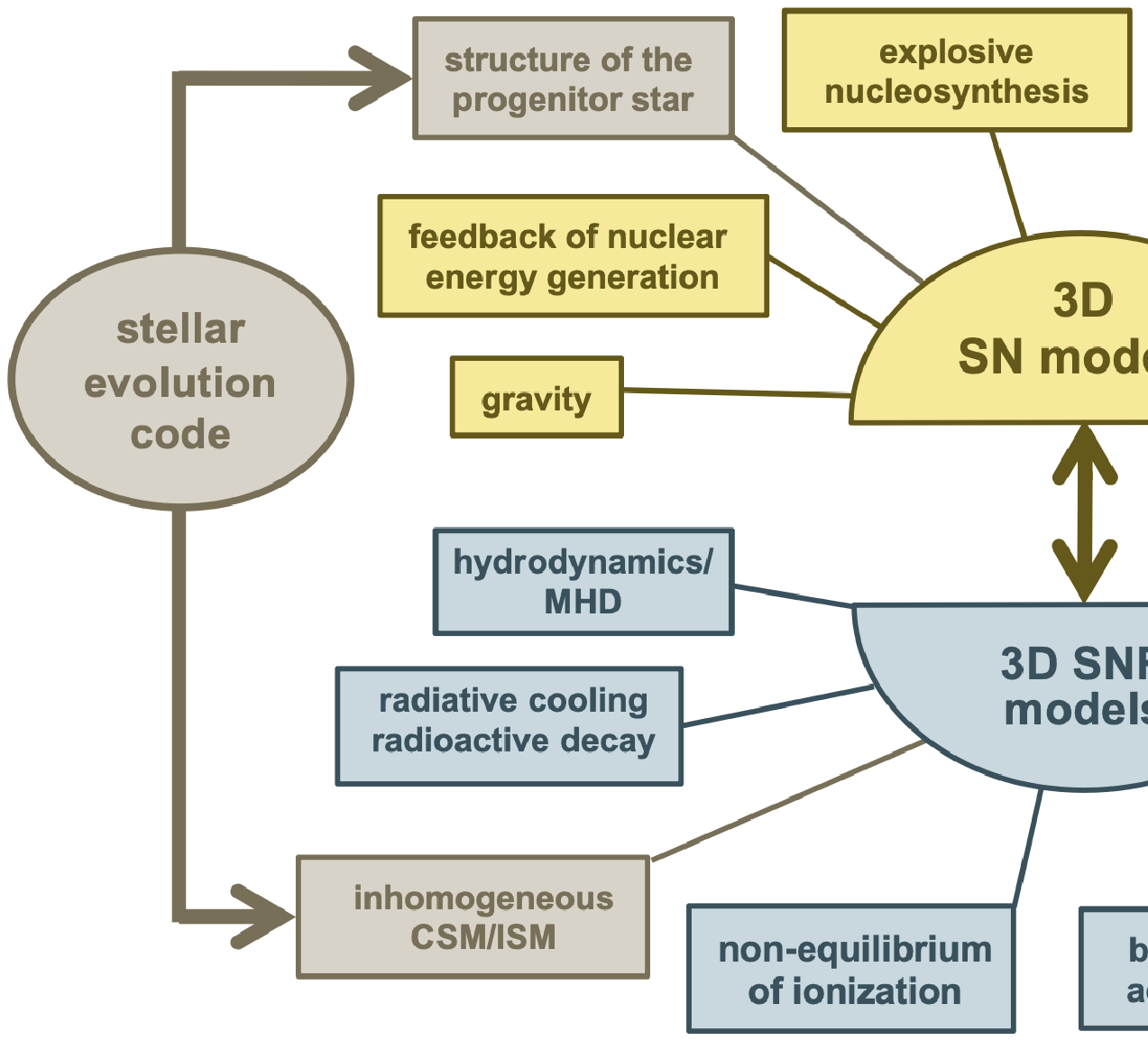}
    \caption{Schematic representation of the strategy adopted to
    model the evolution from the SN to the SNR and to compare
    the model results with multi-wavelength observations.}
  \label{fig1}
  \end{center}
\end{figure}

The approach we have adopted is multi-disciplinary, encompassing
two main aspects: a) the development and use of comprehensive models
that describe the evolution of massive stars, the dynamics of SN
explosions, and the expansion of SNRs; b) the accurate analysis of
multi-band observations of progenitor systems and SNRs, followed
by a detailed comparison of the observational data with the results
obtained from our models. In our modeling efforts, we devised a
comprehensive descriptive framework resulting through the coupling
of state-of-the-art models of stellar evolution, core-collapse SN
and SNR (see Fig.~\ref{fig1}).

In general, we considered stellar models that describe the evolution
of massive stars from the pre-main-sequence phase up to the pre-SN
stage. Throughout our investigations, we have examined diverse
progenitor types, including red supergiants \cite{2002RvMP...74.1015W,
2016ApJ...821...38S}, stripped-envelope supergiants \cite{2017ApJ...842...13W}
(namely red supergiants that have lost their hydrogen envelope
thousands of years before collapse), blue supergiants
\cite{1988PhR...163...13N, 1990ApJ...360..242S}, supergiants resulting
from the merging of two massive stars \cite{2018MNRAS.473L.101U},
and luminous blue variables \cite{2018ApJS..237...13L}. These models
describe the star's configuration at the time of core
collapse, providing the initial conditions for our SN models in the
form of radial profiles of quantities such as density, pressure,
velocity (including the rotational velocity of the star), and various
chemical species distributions.

The 3D SN models (such as those presented in \cite{2017ApJ...842...13W}
and \cite{2020ApJ...888..111O}) describe the intricate phases
subsequent to the core-collapse of the progenitor star. These models
include all the necessary components to elucidate the processes
occurring during these phases (such as radioactive decay, explosive
nucleosynthesis, gravity, and the feedback from nuclear energy
generation) as well as stochastic processes (such as convective overturn
due to neutrino heating and the activity of the standing accretion
shock instability - SASI) that arise and influence the post-explosion
asymmetry in the distribution of ejected materials. The SN models
provide the initial conditions for our SNR simulations a few hours
(typically 1 day) after the core-collapse and soon after the
breakout of the SN blast at the surface of the star.

Our 3D SNR models include all the necessary ingredients to describe
the expansion of the remnant within a non-uniform environment (CSM
or ISM). To be more specific, our models incorporate: a magnetized
surrounding medium; the energy release stemming from radioactive
decay of isotopes (such as the radioactive decay chain
$^{56}$Ni~$\rightarrow, ^{56}$Co~$\rightarrow, ^{56}$Fe
\cite{2021A&A...645A..66O}); deviations from ionization equilibrium
and electron-proton temperature equilibration \cite{2015ApJ...810..168O};
the back-reaction of accelerated cosmic rays \cite{2012ApJ...749..156O};
radiative losses from optically thin plasma; and magnetic-field-oriented
thermal conduction \cite{2008ApJ...678..274O}. Providing an accurate
and realistic description of the ambient environment through which
the remnant expands is a pivotal aspect of our methodology. To
achieve this, our studies focus on specific SNRs that provide
observational constraints for the structure and density distribution
of the surrounding medium (e.g., SN 1987A, Cas~A, IC 443).
Alternatively, we capitalize on the outputs from stellar evolution
models, which, in addition to detailing the progenitor's structure
at collapse, offer insights into its mass-loss history. From these
data, we are able to formulate a coherent description of the CSM
through dedicated 3D MHD simulations.

An essential step in our methodology is the synthesis of observables,
including both thermal and non-thermal emissions across diverse
spectral ranges, as well as the spatial arrangement and chemical
makeup of ejected materials. In this way we can compare
model outcomes (including synthetic spectra, light
curves, and emission maps) against specific observations, thereby
establishing a two-way exchange of insights. Firstly, we can constrain
the models and fine-tune their accuracy with observations, and
secondly, we can interpret with the models the distinctive signatures
of various phenomena evident in the observations.

It is worth noting that our approach has facilitated the link of
modeling attempts that were previously carried out independently,
each focusing on specific aspects. Prior efforts were often constrained
to singular aspects, such as the stellar evolution, the early phase
of the SN up to a few days, or the interaction of SNRs with the
inhomogeneous CSM or ISM, starting from idealized parametrized initial
conditions (at an age of the remnant of hundreds of yeras). Through
the integration of these diverse modeling components into a unified
framework, our objective is to achieve a holistic and coherent
comprehension of the entire trajectory, spanning from the evolution
of the progenitor star to the subsequent phases of SNR development.

\section{Imprints of progenitor stars and supernovae on the remnant
structure}

In the following sections, I will provide a summary of the principal
outcomes obtained so far as a result of the approach outlined in
Sect.~\ref{method}. Notably, it has enabled us to: 1) identify
the asymmetries of the remnants inherited from the SN explosion,
thereby furnishing crucial insights into the mechanisms underpinning
SN explosions \cite{2016ApJ...822...22O, 2020ApJ...888..111O,
2020A&A...642A..67T, 2021A&A...645A..66O}; 2) identify the
fingerprints of the progenitor stellar systems \cite{2020A&A...636A..22O,
2020ApJ...888..111O}, shedding light on the final stages of stellar
evolution; 3) probe the structure and geometry of the surrounding
CSM or ISM around the SNRs \cite{2015ApJ...810..168O,
2019A&A...622A..73O, 2019NatAs...3..236M, 2022A&A...666A...2O,
2021A&A...654A.167U, 2021A&A...649A..14U, 2023MNRAS.518.6377P},
providing insight on the mass loss-history of the progenitor stars
in the final phases of their evolution.

\subsection{Signatures inherited from the asymmetric SN explosions}
\label{sec_SN}

A remnant that offers the possibility to study the effects of the
SN explosion on the final remnant structure and morphology is Cas~A.
This is the remnant of a stripped envelope SN exploded $\approx
350$ years ago. Observations reveal that this particular remnant
exhibits a notably intricate structure, marked by substantial
large-scale asymmetries and a heterogeneous chemical distribution
of ejecta, which collectively shape its distinctive morphology. On
the other hand, observations also suggest that the morphology and
expansion rate of this remnant are both consistent with its expansion
through the (almost) spherically symmetric wind of the progenitor
star \cite{2014ApJ...789....7L}. These lines of evidence suggest
that the bulk of asymmetries in Cas~A was inherited from the SN
explosion rather than from interaction of the remnant with an
inhomogeneous CSM.

Therefore, Cas~A presents an exceptional opportunity to bridge the
structure of a mature SNR with age of a few hundreds years to the
underlying processes that govern the explosion of a core-collapse
SN and that give rise to an asymmetric ejecta structure
shortly after the collapse. Moreover, Cas~A stands as one of the
most extensively examined SNRs in our Galaxy. Its three-dimensional
structure has been meticulously reconstructed with exceptional
precision, thanks to the analysis of multi-wavelength observations
\cite{2010ApJ...725.2038D, 2013ApJ...772..134M, 2015Sci...347..526M,
2020ApJ...889..144H}. The remnant exhibits three extensive regions
rich in iron, surrounded by ejecta abundant in oxygen, silicon, and
sulfur. Additionally, the remnant's structure features two jet-like
formations (or more precisely sprays of ejecta knot rather than
narrow jets \cite{2006ApJ...645..283F}) that are rich in silicon.
All of these prominent large-scale features are believed to originate
from the pristine structures generated during the SN explosion.

In our initial effort to explore potential origins of the large-scale
asymmetries in Cas~A, we described its evolution from the SN event
to its current age of $\approx 350$~years. We adopted SN models
that describe an initial spherically symmetric explosion, to which
parameterized post-explosion anisotropies were introduced by us
shortly after the shock breakout at the stellar surface
\cite{2016ApJ...822...22O}. Our exploration of the parameter space
included a range of variables related to these initial anisotropies;
this included their density and velocity contrasts relative to the
surrounding ejecta, their size, and their position within the
remnant. The aim was to identify a specific set of parameters capable
of reproducing the masses of shocked silicon, sulfur, and iron, as
inferred from the analysis of X-ray observations. Additionally, we
explored how the post-explosion anisotropies determine the overall
structure and morphology of Cas~A. In this way, we were able to
establish a connection between the main asymmetries and geometric
characteristics observed in the bulk ejecta of Cas~A and the physical
attributes of anisotropies that emerged shortly after the SN
explosion.

\begin{figure}
  \begin{center}
    \includegraphics[width=15cm]{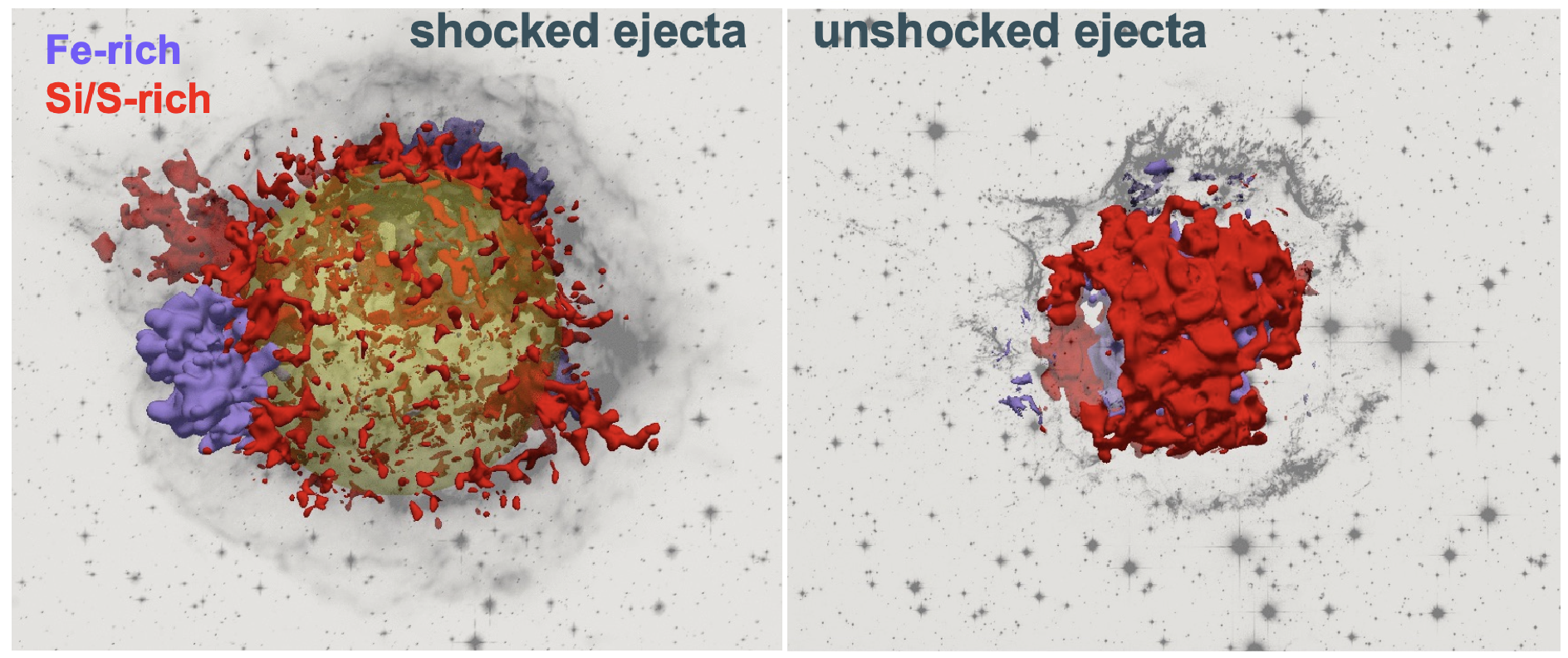}
    \caption{Shocked (on the left) and unshocked (on the right)
    distributions of Fe (blue) and Si/S (red) derived from our
    favorite model of Cas~A at the age of $\sim 350$~years
    \cite{2016ApJ...822...22O}. The yellow sphere on the left panel
    is the average position of the reverse shock. The transparent
    images in the panels are composite Chandra (band [0.3, 10] keV;
    retrieved from www.nasa.gov) and Hubble Space Telescope (image
    sensitive to emission in cold O and S lines; retrieved from
    www.spacetelescope.org) images of Cas~A. A 3D interactive model
    of this simulation can be explored at https://skfb.ly/6RBPs.}
  \label{fig2} \end{center}
\end{figure}

Figure~\ref{fig2} shows the results for our favorite model. The
figure describes the distributions of shocked (on the left) and
unshocked (on the right) Si, S and Fe at the age of Cas~A derived
with our model. The figure also shows actual images of Cas~A obtained
in the optical (Hubble) and X-ray (Chandra) bands to facilitate the
comparison between the modeled distributions of ejecta and the
observed morphology of the remnant. The model was able to capture
the basic properties of the remnant and most of the large-scale
asymmetries that characterize its ejecta distribution: the three
extended Fe-rich regions and the two Si-rich jet-like
structures. In this way, we were able to constrain the average
physical characteristics (in particular the energy and masses)
of post-explosion anisotropies that are able to reproduce the
observed distributions of Si, S, and Fe observed in Cas~A.

The pioneering results obtained with our model can serve as a
valuable reference for exploring self-consistent SN models. Unlike
the SN models discussed above, these self-consistent models incorporate
post-explosion anisotropies that naturally arise due to stochastic
processes, such as convective overturn due to neutrino heating and
the SASI activities, occurring soon after the core-collapse.
State-of-the-art models of this nature are those developed by the
SN group at Max Planck Institute for Astrophysics in Garching. This
group has undertaken an extensive simulation campaign, producing a
series of 3D simulations describing neutrino-driven SN explosions,
and searched for a model that has the potential to reproduce the
fundamental characteristics of Cas~A \cite{2017ApJ...842...13W}.
Remarkably, one of their models produces an ejecta distribution
characterized by three distinct Ni-rich fingers that could correspond
to the extended Fe-rich regions observed in Cas~A. Interestingly,
this model considers, as initial conditions for the SN explosion,
a progenitor star with an initial mass of $15\,M_{\odot}$ that has
lost most of its H-envelope down to a rest of $\approx 0.3\,M_{\odot}$
at the time of collapse \cite{2017ApJ...842...13W}. This aligns
with the evidence from observations of light echoes, which indicated
that Cas~A is the remnant of a type IIb SN (e.g.,
\cite{2008Sci...320.1195K, 2011ApJ...732....3R}). These types of
SN simulations cease shortly after the shock breakout at the stellar
surface, covering only one day of evolution.

\begin{figure}
  \begin{center}
    \includegraphics[width=15cm]{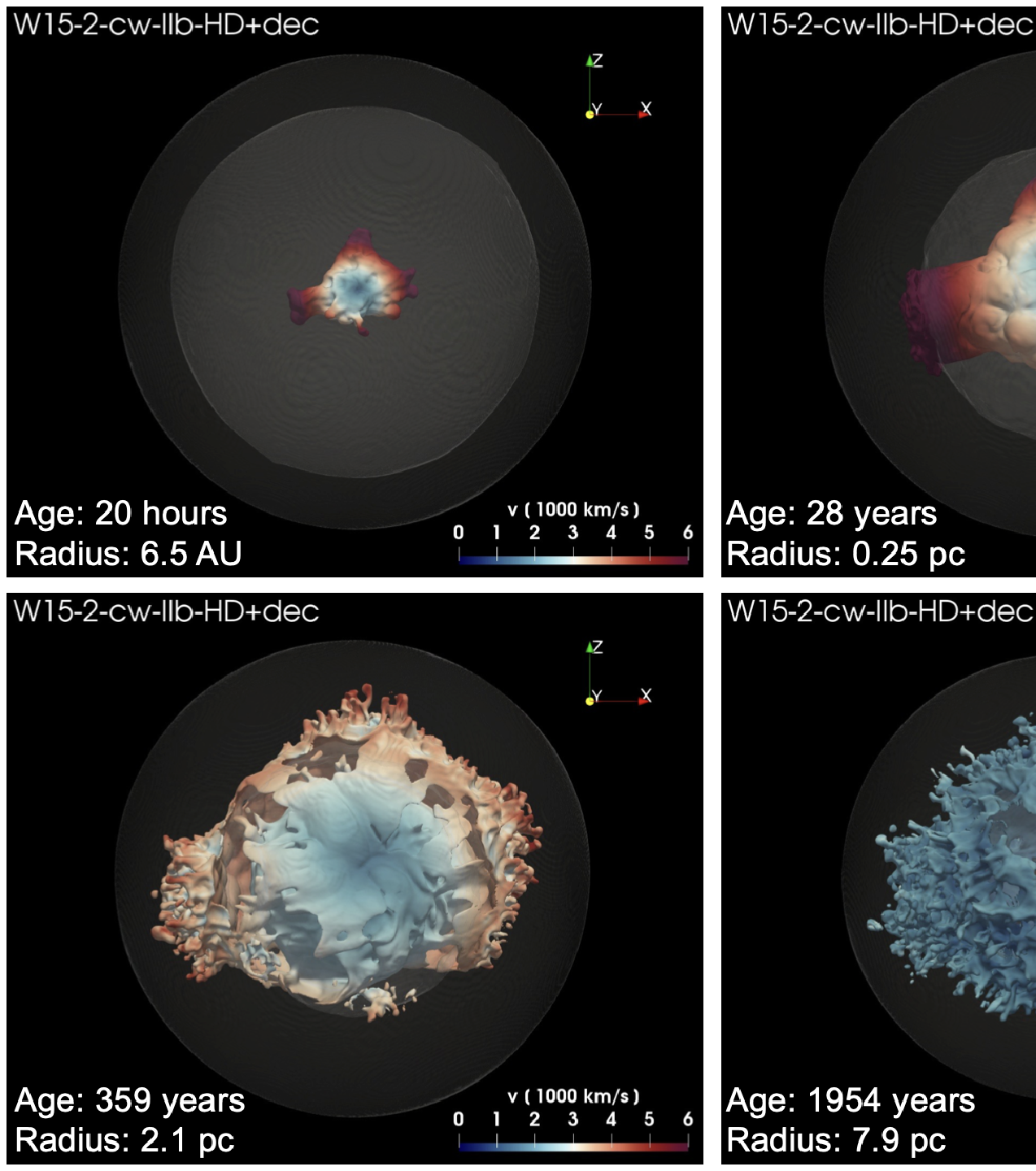}
    \caption{Isosurfaces of the distribution of Ni and Fe for the
    remnant of a neutrino-driven SN (model W15-2-cw-IIb-HD+dec;
    \cite{2021A&A...645A..66O}). The upper left panel shows the
    isosurface of the distribution of Ni soon after the shock
    breakout at the stellar surface, $\approx 20$ hours after the
    core collapse. The other panels show the isosurface of the
    distribution of Fe (the final product of the radioactive decay
    chain $^{56}$Ni~$\rightarrow, ^{56}$Co~$\rightarrow, ^{56}$Fe)
    at the time when the reverse shock starts to interact with the
    Fe-rich plumes of ejecta (upper right panel), at the age of
    Cas~A (lower left), and at $t\approx 2000$~years (lower
    right). The opaque irregular isosurfaces correspond to a
    value of Ni or Fe density which is at 5\% of the peak density;
    their colors give the radial velocity in units of 1000 km
    s$^{-1}$ on the isosurface (the color coding is defined at the
    bottom of each panel). The semi-transparent quasi-spherical
    surfaces indicate the forward (internal sphere) and reverse
    (external sphere) shocks. A 3D interactive model of this
    simulation can be explored at https://skfb.ly/6TKRK.}
  \label{fig3} \end{center}
\end{figure}

As a follow-up to the above study, we have extended the most promising
SN simulation conducted by the Garching group. This extension aims
to comprehensively describe the subsequent evolution of the SNR
over a span of thousands of years \cite{2021A&A...645A..66O}.
Notably, this simulation represents the first and currently the
only one that comprehensively describes the entire 3D evolution of
a neutrino-driven SN explosion, spanning from the core-collapse
phase to the SNR's mature age of 2000 years (see Fig.~\ref{fig3}).
We found that immediately after the shock breakout, both the blast
wave and the ejecta freely expand through the spherically symmetric
wind of the progenitor star. During the initial year of evolution,
the locally deposited energy from the radioactive decay chain
$^{56}$Ni~$\rightarrow, ^{56}$Co~$\rightarrow, ^{56}$Fe results in
inflation of the initially Ni-rich ejecta. This phenomenon is
commonly referred to as the "Ni-bubble effect" (see also
\cite{2021MNRAS.502.3264G}). Subsequently, the Fe-rich plumes expand
and start the interaction with the reverse shock approximately 30
years after the SN event. This is the time when the Fe-rich regions,
observed in Cas~A, begin to form as a result of this interaction.

In the subsequent stages of evolution, the expansion of the ejecta
continues. HD instabilities manifest at the contact discontinuity,
giving rise to the formation of ring and crown structures within
the shocked Fe-rich regions when the SNR is approximately 350 years
old. These smaller-scale features bear a qualitative resemblance
to those observed in Cas~A \cite{2013ApJ...772..134M}. As for the
unshocked ejecta, the simulation reveals that voids and cavities
constitute their structural characteristics, mirroring what is
observed in Cas~A \cite{2015Sci...347..526M}. In our model, these
cavities result from the efficient Ni-bubble effect during the first
year of evolution. Furthermore, the simulation illustrates that the
three Fe-rich plumes inherited from the SN define a plane where the
ejecta, rich in Ti and Fe, expand preferentially. This configuration
gives rise to a "thick-disk" geometry for the ejecta, which is
tilted at an angle of approximately -30 degrees relative to the
plane of the sky. This geometry (and orientation) bears a striking
resemblance to the one observed in Cas~A \cite{2010ApJ...725.2038D,
2013ApJ...772..134M}. In the subsequent phases of the evolution,
the model describes the continued expansion of the remnant until
it reaches an age of 2000 years. This comprehensive approach allowed
us to make predictions about the future evolution of Cas~A and compare
the model results with those of other observed SNRs. For example,
we compared our findings with the Galactic core-collapse SNR
G292.0+1.8, an older counterpart to Cas~A with an estimated age of
approximately 2000 years \cite{2022ApJ...932..117L}.

We note that our neutrino-driven SNR model does not account for the
Si-rich jets that are observable in Cas~A. We proposed that these
jets might originate from a post-explosion phenomenon not included
in our simulations. For example, the presence of an accretion disk
composed of fallback matter around the newly formed neutron star
could potentially give rise to the jets through a mechanism similar
to that responsible for the formation of jets in pulsars
\cite{2021A&A...645A..66O}.

Nonetheless, our simulations have effectively demonstrated that a
neutrino-driven SN explosion provides a natural explanation for the
majority of the observed asymmetries and features within Cas~A.
Specifically, the observed structure of the ejecta arises from the
interaction between the reverse shock and the initial large-scale
asymmetries that emerged due to stochastic processes during the SN.
These processes, such as convective overturn and SASI activities,
initiated within the first few seconds of the SN explosion. In
essence, our findings establish that even at an age of hundreds of
years, SNRs still retain traces of the events that occurred immediately
following the core-collapse. As a result, they carry valuable
information about the physics of the SN engine.

\subsection{The imprint of the progenitor star}
\label{sec_star}

SNRs are also expected to potentially preserve traces of their
progenitor star's structure. However, discerning the progenitor's
influence on the final remnant structure is a complex task due to
the necessity of untangling the effects of the progenitor from those
arising from the SN explosion and the interaction of the remnant
with the CSM. Among historical SNRs, that of SN 1987A presents
the most promising prospects for identifying the progenitor's imprint
on the SNR structure. This is primarily because the evolution of
the SN and the subsequent transition to the SNR phase have been
continuously monitored across various wavelengths. Moreover,
observations of the progenitor star were collected prior to its
collapse, providing us with comprehensive details regarding its
properties, including its luminosity and photospheric temperature
(i.e., its position in the Hertzsprung–Russell diagram). These
factors collectively establish SN 1987A as the prime target for
investigating the intricate connection between the progenitor star,
the SN, and the resulting SNR.

Like many SNRs, SNR 1987A exhibits multiple asymmetries within its
morphology. The most prominent ones, such as the ring-like structure
visible across all wavelengths, are primarily attributed to the
interaction of the remnant with a complex and dense CSM (see Sect.
\ref{sec_env}). Our current focus centers on the anisotropies
characterizing the structure of the unshocked ejecta, specifically
the stellar debris that has not yet interacted with the reverse
shock and that, therefore, is not affected by the interaction with
the CSM. As a result, this portion of the ejecta has the potential
to retain both the characteristics of the progenitor star's structure
and the hallmarks of the SN explosion.

Observations made 1-2 years after the SN event have unveiled iron
lines displaying a redshifted centroid at approximately $\approx
280 \pm 140$~km~s$^{-1}$, with wings extending beyond $> 3000$~km~s$^{-1}$
\cite{1990ApJ...360..257H}. These findings suggest the presence
of metal-rich ejecta forming clumps rapidly receding from the
observer. Around 20 years later, NuSTAR observations corroborated
this discovery by detecting lines resulting from the decay of
$^{44}$Ti, once again revealing a redshifted Doppler velocity of
approximately $\approx 700$~km~s$^{-1}$ \cite{2015Sci...348..670B}.
This reaffirms the presence of metal-rich ejecta moving away from
the observer at velocities in the hundreds of km~s$^{-1}$ range.
Finally, recent ALMA observations has allowed researchers to delve
into the innermost structure of the unshocked ejecta. These
observations facilitated the reconstruction of the 3D distribution
of molecules such as carbon monoxide (CO) and silicon monoxide (SiO)
\cite{2017ApJ...842L..24A}. Intriguingly, they demonstrated that
these molecules are arranged in a torus-like distribution tilted
at an angle of about $90^{\rm o}$ with respect to the dense inner
equatorial ring that characterize the CSM \cite{2005ApJS..159...60S}.

All of the above lines of evidence collectively and robustly indicate
that the explosion giving rise to SN 1987A was profoundly asymmetric
in nature. Given that the development of post-explosion asymmetries
(due to stochastic processes following the collapse) can reflect
the internal structure of the stellar interior where they originate
\cite{2015A&A...577A..48W}, a natural question is: could insights
into the structure of the progenitor star at the time of its collapse,
and consequently into its nature, be obtained from the analysis of
these asymmetries?

To answer the above question, we coupled together stellar evolution
models, 3D core-collapse SN models, and 3D SNR models
\cite{2020A&A...636A..22O, 2020ApJ...888..111O}. The stellar models
describe the evolution of the progenitor star and its structure at
the time of collapse. We explored three scenarios: a blue supergiant
that evolved as a solitary star \cite{1988PhR...163...13N}, a blue
supergiant formed through the slow-merger of two massive stars
\cite{2018MNRAS.473L.101U}, and a red supergiant
\cite{2002RvMP...74.1015W}. Of particular interest was the slow-merger
scenario, as it proved capable of reproducing a majority of the
observational constraints linked to the progenitor star of SN 1987A
(Sanduleak $-69^{\rm o}, 202$). This model successfully replicated the
star's transition from red to blue, its lifespan of approximately
20,000 years, its total mass at the time of collapse (around $18.3
M_{\odot}$), and its position in the Hertzsprung-Russell diagram
(with values of $\log T_{\rm eff} = 4.2$ and $\log L/L_{\odot} =
4.9$) at the time of collapse. To further explore the sensitivity
of a remnant to the progenitor's structure, we also considered a
red supergiant model, which exhibited a substantially different
stellar structure compared to the other two models.

The progenitor models served as the basis for establishing the
initial conditions in our 3D SN simulations. For our purposes, we
characterized the initial asymmetry of the SN by assuming a bipolar
explosion. This asymmetry was regulated by two key parameters: the
ratio of energy deposition along the polar axis in comparison to
the equatorial plane at the boundary of the collapsing core, and
the ratio between the energy deposition at the two polar regions.
Then, we explored the parameter space searching for an optimal set
of SN parameters and the orientation of the bipolar explosion with
respect to the line of sight that is able to reproduce the observations
of SN 1987A as closest as possible. Specifically, this exploration
aimed to closely replicate the profiles of Fe lines observed
approximately 1-2 years following the SN event. Our SN models covered
the initial 20 hours of evolution, and their results provided the
initial conditions for the subsequent SNR simulations.

In the final phase of our modeling, we conducted 3D SNR simulations,
describing the interaction of the SN blast wave with the non-uniform
CSM. This CSM consists of an extensive H\,II region, creating a
torus-like dense environment in close proximity to the explosion's
center, and a dense equatorial ring immersed within the H\,II region.
As with the SN models, we systematically explored the parameter
space of the SNR models. Our objective was to identify an optimal
set of parameters that would faithfully replicate the X-ray emissions
observed from SNR 1987A.

\begin{figure}
  \begin{center}
    \includegraphics[width=15cm]{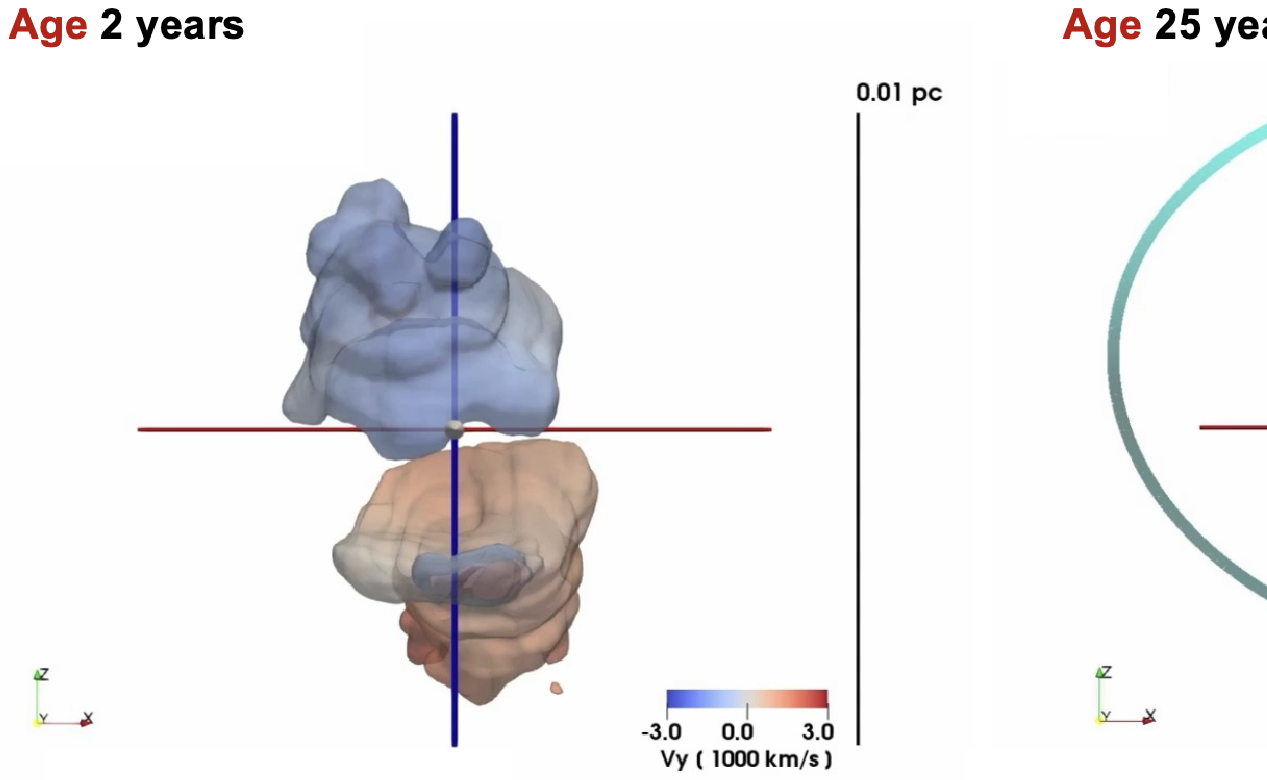}
    \caption{Isosurfaces for the distributions of Fe (left panel),
    Ti (center panel), and the CO and SiO molecules (right panel)
    at the labeled times for our preferred model of SN 1987A
    (B18.3; adapted from \cite{2020A&A...636A..22O}). The colors on the
    isosurfaces in the left and center panels give the velocity
    along the line of sight in units of 1000~km~s$^{-1}$ on the isosurface
    (the color coding is defined at the bottom of each panel). The
    yardstick indicating the length scale is on the left of these
    two panels. The ring in the center and right panels indicates
    the position of the dense equatorial ring. The colors of the
    isosufaces displayed in the right panel mark the distributions
    of CO (red) and SiO (green). A 3D interactive model of this
    simulation can be explored at https://skfb.ly/6RGKP.}
  \label{fig4} \end{center}
\end{figure}

Using our models, we generated observables of the SNR that include
emission maps, spectra, light curves, and spatial distributions of
ejecta and molecules. We then compared these model-derived results
with actual observations collected throughout the entire evolution,
spanning from the SN phase to the present age of the fully-developed
SNR. Our analysis revealed that the model that most accurately
replicates the majority of observables corresponds to the slow-merger
scenario as the progenitor stellar system. Figure~\ref{fig4} displays
selected outcomes from our best-fit model. Notably, this model
predicts the presence of two Fe-rich clumps moving in opposite
directions: a heavier one moving southward, away from the observer,
and the other heading northward, toward the observer (see left panel
of Fig.~\ref{fig4}). These clumps lead to Fe line profiles at the
age of 1-2 years that strikingly resemble the observed profiles.
Specifically, the lines are redshifted, with a peak velocity of
approximately 300~km~s$^{-1}$, and exhibit broadening, with wings
extending to velocities exceeding 3000~km~s$^{-1}$.
Additionally, our model enabled us to constrain the energy of the
explosion, estimated to be around $\sim 2\times 10^{51}$~erg, and
the orientation of the bipolar explosion relative to the line of
sight.

At the age of 25 years, our analysis revealed that the innermost
structure of the ejecta still exhibits the presence of two clumps,
both rich in Ti and Fe (see center panel of Fig.~\ref{fig4}). We
generated synthetic lines from the decay of $^{44}$Ti and compared
them with observations collected by NuSTAR at the same age. Once
again, we determined that the model providing the closest match to
the observed lines (reproducing its redshift and broadening)
corresponds to the scenario in which the progenitor star resulted
from a merger. The redshift observed in the Fe and Ti lines primarily
stems from the inherent asymmetry in the energy deposition, as
assumed in the bipolar SN model, at the two polar regions.
Interestingly, our model also reproduces many of the properties of
the two Fe-rich clumps recently revealed by JWST observations
\cite{2023ApJ...949L..27L}.

Finally, we generated spatial distributions of CO and SiO molecules
at the age of 25-27 years, as illustrated in the right panel of
Fig.~\ref{fig4}. Then, we compared these outcomes with distributions
derived from ALMA observations. We found that the observed distributions
align quite closely with models assuming a blue supergiant as the
progenitor star. Notably, the model based on a red supergiant
progenitor fails to replicate these distributions. This implies
that the structure of the ejecta exhibits sensitivity to the
progenitor star's structure at the time of collapse, keeping memory
of the characteristics of the progenitor even decades after the SN
explosion. Indeed both the other two models (assuming a blue supergiant
progenitor) successfully reproduced the toroidal distribution of
the molecules. It's noteworthy that these models predict a size and
orientation in the space for the torus-like distributions of CO and
SiO that remarkably align with those observed. Consequently, we
concluded that this particular feature serves not only as a signature
of the progenitor star but also as an indicator of the bipolar
explosion and its orientation in space.

\subsection{Effects of the circumstellar medium on the remnant morphology}
\label{sec_env}

As previously discussed, the most notable features of SNR 1987A are
a direct consequence of its interaction with a highly inhomogeneous
CSM. Detailed observations across various wavelengths, including
radio, infrared, optical, and X-ray bands, show a remnant distinguished
by a prominent ring-like structure. This distinctive morphology
arises from the collision between the SN blast and an H\,II region,
as well as a dense equatorial ring, around the center of explosion
\cite{2005ApJS..159...60S}. Constraining the density distribution
of the CSM is important to gain insights into the mass loss history
of the progenitor star during the concluding stages of its evolution.

We used our model, which describes the evolution from SN 1987A to
its remnant (as outlined in Sect.~\ref{sec_star}), to constrain the
geometry and density distribution of the surrounding CSM. Additionally,
we assessed the contributions of various plasma components to both
the thermal and non-thermal emissions originating from the remnant
\cite{2015ApJ...810..168O, 2019A&A...622A..73O,
2020A&A...636A..22O}. This achievement was made possible through
a meticulous comparison of the model's outcomes with radio and X-ray
light curves and emission maps \cite{2019A&A...622A..73O,
2020A&A...636A..22O}, as well as X-ray spectra \cite{2019NatAs...3..236M},
all derived from actual observations spanning approximately 30 years
of evolution.

\begin{figure}
  \begin{center}
    \includegraphics[width=15cm]{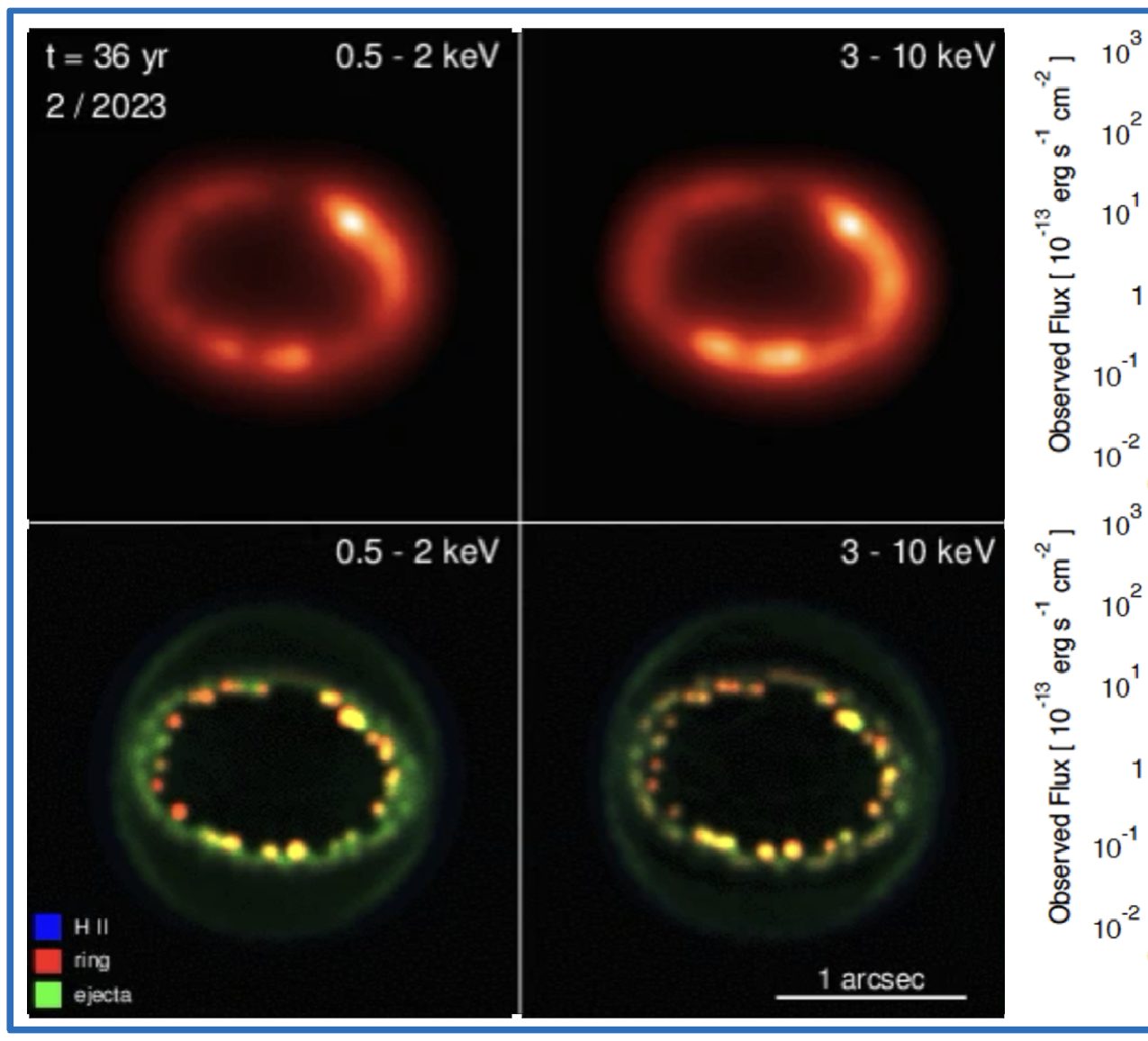}
    \caption{Results for our preferred model of SNR 1987A (B18.3)
    illustrating the interaction of the remnant with the inhomogeneous
    CSM (adapted from \cite{2020A&A...636A..22O}). Left and center
    panels: Synthetic X-ray emission maps from the model at the
    remnant age of 36~years. The panels display the emission maps
    in the $[0.5,2]$~keV range (upper left) and $[3,10]$~keV range
    (upper right), along with their corresponding three-color
    composite representations (lower left and right). Each image
    is normalized to its maximum for better visibility. The X-ray
    maps were convolved with a Gaussian of size 0.15 arcsec to
    approximate the spatial resolution of {\it Chandra}, while the
    three-color composite images were smoothed with a Gaussian of
    0.025 arcsec. The colors in the composite images represent the
    emission from shocked ejecta (green), shocked plasma from the
    ring (red), and the H\,II region (blue). A 3D interactive model
    of this simulation can be explored at https://skfb.ly/6QTXE.
    Right panels: X-ray light curves (solid lines) in the $[0.5,
    2]$~keV range (upper panel) and $[3, 10]$~keV range (lower
    panel), synthesized from the model and compared to the light
    curves of SN\,1987A observed with various instruments (legend
    located in the upper left corner of the upper panel). Dotted,
    dashed, and dot-dashed lines denote the contributions to the
    emission from the shocked ejecta, shocked plasma from the H\,II
    region, and shocked plasma from the ring, respectively. The
    vertical red dash-dotted lines indicate the timing of the
    emission maps shown on the left.}
  \label{fig5} \end{center}
\end{figure}

As an illustrative example, Fig.~\ref{fig5} displays the X-ray
emission synthesized by our best-fit model of SN 1987A and the
comparison with observational data \cite{2020A&A...636A..22O}. We
found that the same model, which reproduces the asymmetries in the
distribution of ejecta inherited from the SN explosion (see
Sect.~\ref{sec_star}), also successfully reproduces the asymmetries
induced by the interaction of the remnant with the inhomogeneous
CSM, leading to the observed X-ray morphology (emission maps on the
left panels of the figure) and the light curves in both the soft
and hard X-ray bands (displayed in the panels on the right of the
figure). We also found that the same model is able to replicate
remarkably well high- and low-resolution X-ray spectra collected
for SN 1987A in different stages of evolution \cite{2019NatAs...3..236M}
as well as radio maps and light curve \cite{2019A&A...622A..73O}.

According to our model, the interaction between the remnant and the
CSM started approximately three years after the SN event. During
this phase, the shocked material from the H\,II region and the outer
shocked ejecta caused a rapid increase in X-ray emission, as evident
in the light curves. Simultaneously, it led to the distinctive
ring-like morphology observed in the remnant. This phase persisted
for roughly 14 years. Subsequently, the soft light curve exhibited
a further steepening when the blast wave encountered the dense
equatorial ring. In this second phase of evolution, the X-ray
emission was predominantly driven by the shocked material from the
ring. According to our model, this phase concluded around 34 years
after the SN explosion when the soft X-ray light curve significantly
flattened and, then, slightly decreased. In the subsequent
phase of evolution, the X-ray emission began to be predominantly
driven by the shocked ejecta.

In the case of SNR 1987A, the influence of the CSM on the structure
and appearance of the remnant is clearly evident. Consequently, it
is not surprising that the comparison between our models and
observations has enabled us to constrain the CSM's geometry and
density distribution, as well as offer predictions on the future
evolution of the remnant. In the next case, I illustrate how our
approach can be very useful, even when there is no apparent evidence
of an interaction between the remnant and an inhomogeneous CSM in
the data.

\begin{figure}
  \begin{center}
    \includegraphics[width=15cm]{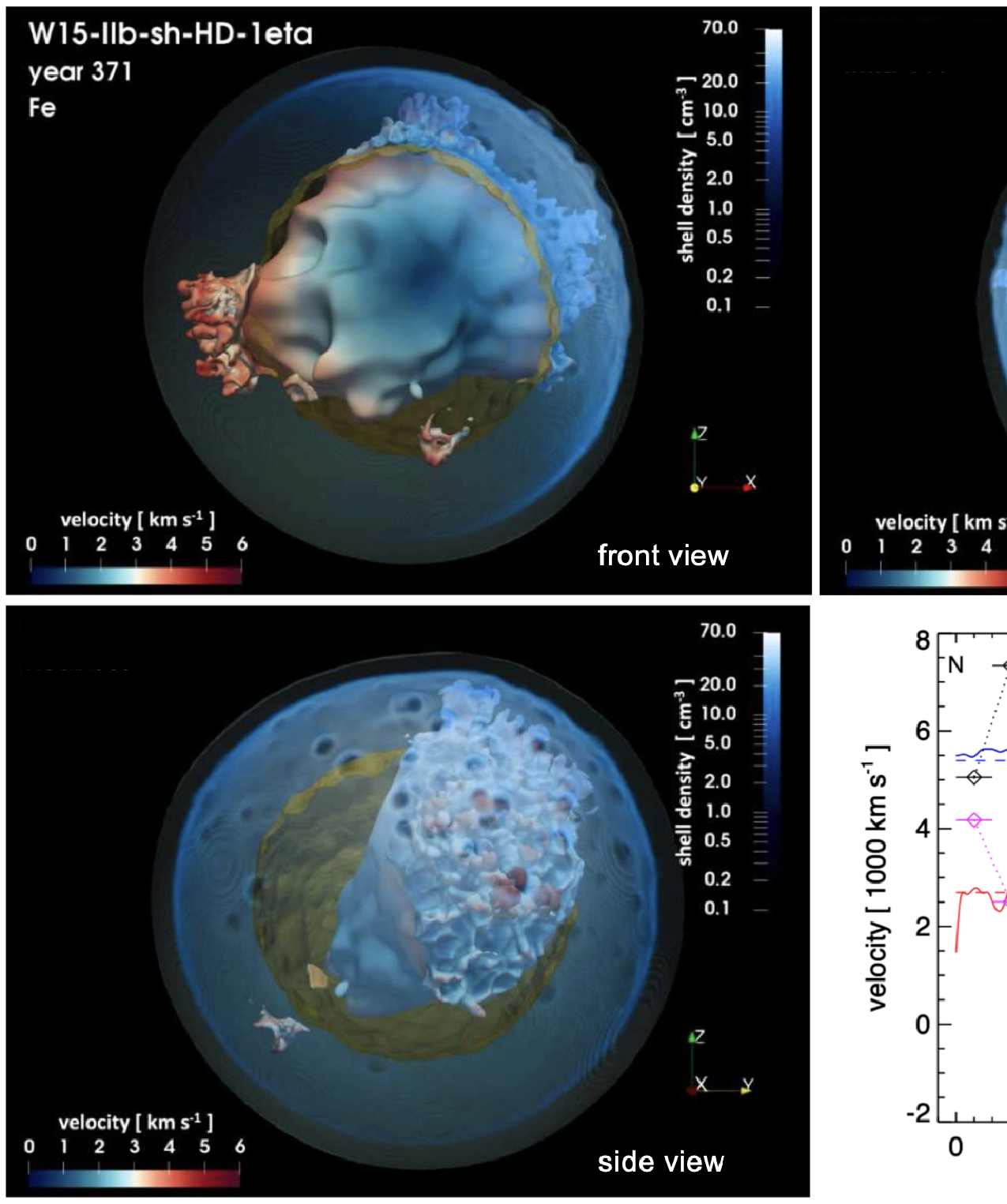}
    \caption{Results for one of the models (W15-IIb-sh-HD-1eta)
    describing the interaction of Cas~A with an asymmetric circumstellar
    shell (adapted from \cite{2022A&A...666A...2O}). Upper panels
    and lower left panel: Isosurfaces displaying the distribution
    of iron at the age of Cas~A from various viewing angles. Colors
    represent radial velocity in units of 1000~km~s$^{-1}$ (color
    coding is explained at the bottom of each panel). The
    semi-transparent clipped quasi-spherical surfaces indicate the
    forward (green) and reverse (yellow) shocks. The shocked shell
    is visualized using volume rendering with a blue color palette
    (color code on the right of each panel), and opacity corresponds
    to plasma density. A 3D interactive model of this simulation
    can be explored at https://skfb.ly/o8FnO.  Lower right panel:
    Forward shock velocities (in blue) and reverse shock velocities
    (in red) as a function of the position angle in the plane of
    the sky at the age of Cas~A. Dashed horizontal lines denote the
    median values of these velocities. Velocities of the forward
    (represented by black diamonds) and reverse (represented by
    magenta diamonds) shocks, derived from the analysis of Chandra
    observations \cite{2022ApJ...929...57V}, are overlaid for
    comparison.}
  \label{fig6} \end{center}
\end{figure}

In Sect.\ref{sec_SN}, it is mentioned that observations suggest
that SNR Cas~A is expanding through the spherically symmetric wind
of its progenitor star. However, some asymmetries of its reverse
shock cannot be explained if the CSM around the remnant is spherically
symmetric. These include the inward-moving reverse shock observed
in the western hemisphere \cite{2022ApJ...929...57V}, the offset
between the geometric centers of the reverse and forward shocks
\cite{2001ApJ...552L..39G}, and the evidence that nonthermal emission
from the reverse shock is brighter in the western region compared
to the eastern region \cite{2008ApJ...686.1094H}. In fact, models
of Cas~A that assume a spherically symmetric CSM, in which the
remnant expands, fail to replicate these asymmetries \cite{2021A&A...645A..66O}.

Motivated by the discrepancies evidenced between models and
observations, we explored the possibility that Cas A encountered
an inhomogeneous CSM during its expansion. More precisely, we
postulated that the remnant interacted with an asymmetric circumstellar
shell in the past. We then explored the parameter space of this shell,
searching for a configuration that could account for the observed
asymmetries in the reverse shock of Cas~A \cite{2022A&A...666A...2O}.

We have found that the interaction between the remnant and a
circumstellar shell can generate asymmetries resembling those
observed in the reverse shock. Figure~\ref{fig6} illustrates the
results of our favourite model in which the asymmetric shell had
its densest portion located on the nearside to the northwest.
According to this model, the shell was relatively thin, with a
thickness of approximately $\sigma \approx 0.02$~pc, and its radius
extended to about $1.5$~pc from the center of the explosion. According
to the model, at the current age of Cas~A, the reverse shock exhibits
several notable asymmetries which resamble those of Cas~A: it moves
inward in the observer's frame within the northwest region while
moving outward in most other regions; the geometric center of the
reverse shock is offset to the northwest by approximately $\approx
0.1$~pc from the geometric center of the forward shock; and the
reverse shock in the northwest region displays enhanced nonthermal
emission because the ejecta enter the reverse shock there with a
higher relative velocity (ranging between $4000$ and $7000$~km~s$^{-1}$)
compared to other regions (below $2000$~km~s$^{-1}$).

We concluded that the large-scale asymmetries observed in Cas~A's
reverse shock can be attributed to the interaction between the
remnant and an asymmetric, dense circumstellar shell that occurred
between approximately $180$ and $240$~years after the SN event.
This shell was likely the result of a massive eruption (with an
estimated mass on the order of $2\,M_{\odot}$) from the progenitor
star, which took place between $10^4$ and $10^5$~years before its
core collapse.

\section{Summary and Conclusions}

In this brief review, I have presented an approach that we have
developed to follow the path from massive stars to SNe and SNRs in
a coherent way. This approach enables us to explore the connection
between progenitor stars, SNe, and SNRs. Indeed there is a broad
consensus in the literature that the morphologies and properties
of SNRs reflect various factors, including: asymmetries inherited
from the parent SN explosion; the structure of the progenitor stars
at collapse; the interaction of the remnants with the inhomogeneous
ambient environment (CSM or ISM). Therefore, deciphering multi-wavelength
observations of SNRs can be of paramount importance for extracting
valuable information about: the complex phases in SN evolution
following core-collapse, shedding light on the processes governing
the SN engine; the internal structure of the progenitor stars,
providing insights into their nature; the structure of the CSM,
offering clues about the mass-loss history of the progenitor stellar
systems. This underscores why the study of the progenitor-SN-SNR 
connection has breakthrough potential to open new exploring windows
on the physics of massive stars, SNe, and SNRs.

Our approach is based on the development of 3D HD/MHD models that
describe the evolution from the core-collapse SN to the fully
developed SNR. This approach incorporates information regarding the
structure of the progenitor star and the CSM, derived from stellar
evolution models and observational analyses. In this review, I have
highlighted a few illustrative applications of this methodology,
with a specific focus on the cases of Cas~A and SN 1987A. These
applications demonstrate how we can extract valuable insights from
the observations of these SNRs, providing us with information about
the physics of the SN engine, the nature of the progenitor star,
and the characteristics of the CSM.

Furthermore, our approach can also be highly effective in investigating
the "after-life" of massive stars, particularly concerning the
formation of compact objects situated at the centers of SNRs. While
this aspect was not the focus of the present review, it is worth
mentioning the case of SN 1987A. Indeed, in this case, the comparison
of our SN 1987A model with observations evidenced the presence of
an excess in the hard X-ray band at energies above 10~keV
\cite{2021ApJ...908L..45G}. Our subsequent analysis allowed us to
identify this excess with a non-thermal source compatible with the
presence of a pulsar wind nebula surrounded by the dense and cold
ejecta material at the heart of the remnant \cite{2021ApJ...908L..45G,
2022ApJ...931..132G, 2023ApJ...949...97D}.

We have determined that the approach outlined here can be very
useful and effective in investigating the progenitor-SN-SNR connection.
However, it's important to note that deciphering the observations
may critically depend on the models. Consequently, these models
should establish a self-consistent link between progenitor stars,
SNe, and SNRs, necessitating a multi-physics, multi-scale, and
multi-dimensional approach. Furthermore, these models should be
firmly based on solid observational data, accounting for the
dynamics, energetics, and spectral properties of SNe and SNRs. As
a result, the synthesis of observables from the models, including
spectra, emission maps, light curves, and more, becomes of paramount
importance in order to rigorously constrain the models.

These models can also serve as powerful tools for the analysis and
interpretation of data. This is of particular significance in the
current era and in the near future, given the wealth of high-quality
data that will be gathered. The observational datasets obtained
with state-of-the-art instruments like JWST and XRISM, as well as
the forthcoming contributions from facilities such as SKA, CTA,
LEM, and Athena, will unquestionably require advanced tools for
their analysis and interpretation. In this context, the models we
have discussed here can play a crucial role as indispensable tools,
positioned to bridge the gap between observational data and meaningful
interpretation.

\acknowledgments
I thank the referee, Hiroyuki Uchida, for the useful feedback that
allowed me to improve the paper. Many colleagues have contributed
to the findings reported here; in particular, I am grateful to F.
Bocchino, A. Dohi, E. Greco, H.-T. Janka, M. Miceli, S. Nagataki,
M. Ono, O. Petruk, A. Tutone, S. Ustamujic, A. Wongwathanarat for
their fundamental contributions to the success of the project. I
acknowledge partial financial contribution from the PRIN INAF 2019
grant ``From massive stars to supernovae and supernova remnants:
driving mass, energy and cosmic rays in our Galaxy''. The navigable
3D graphics have been developed in the framework of the project
3DMAP-VR (3-Dimensional Modeling of Astrophysical Phenomena in
Virtual Reality; \cite{2019RNAAS...3..176O, 2023MmSAI..94a..13O})
at INAF-Osservatorio Astronomico di Palermo.

\bibliographystyle{aa}
\bibliography{references}

%

\bigskip
\bigskip
\noindent {\bf DISCUSSION}

\bigskip
\noindent {\bf JOHN BALLY:} How do you explain the two outer-rings
in SN 1987A? These are the rings displaced above and below the plane
of the inner, X-ray ring.

\bigskip
\noindent {\bf SALVATORE ORLANDO:} The origin of the outer rings
in SN 1987A was well explained by Thomas Morris and Philipp
Podsiadlowski in 2007 (Morris \& Podsiadlowski 2007, Science 315,
pp. 1103). These authors found that the mass ejection resulting
from the merger between two massive stars (an event that occurred
approximately 20,000 years before the supernova explosion) can
accurately replicate the characteristics of the triple-ring nebula
encircling the supernova. In our simulations of SN 1987A, we have
also incorporated the presence of the outer rings in the description
of the CSM. However, our findings revealed that the X-ray emission
stemming from the shocked outer rings is orders of magnitude lower
than that originating from the inner ring. This difference explains
why their X-ray emission has not been detected yet.

\end{document}